\documentclass[preprint,showpacs,showkeys,amsmath,amssymb]{revtex4}

\begin{document}

\title{Inhomogeneus Inflation and Cosmic no-Hair Conjecture}

\vspace{.2in}
\author {M. A. S. Nobre$^{1,2}$\footnote[2]{assunta@fisica.ufpb.br}, M. R. de Garcia Maia$^{1}$\footnote[1]{mrgm@dfte.ufrn.br},
J. C. Carvalho$^{2}$\footnote[3]{carvalho@dfte.ufrn.br}, J. A. S. Lima$^{1,3}$\footnote[4]{limajas@astro.iag.usp.br}}

\smallskip
\address{$^1$Departamento de F\'{\i}sica Te\'orica e Experimental, \\
Universidade Federal do Rio Grande do Norte,
59072 - 970, Natal, RN, Brazil}
\address{$^2$Departamento de F\'{\i}sica, Universidade Federal da Para\'{\i}ba,\\
58051-970, J. Pessoa, PB, Brazil}
\address{$^3$Departamento de Astronomia, Universidade de S\~ao
Paulo, \\
Rua do Mat\~ao, 1226, CEP 05508-900, S\~ao Paulo,
SP, Brazil}

\date{\today}

\begin{abstract}
\noindent
The cosmic no hair conjecture is tested for a large class of inhomogeneous cosmologies
with a positive cosmological constant $\Lambda$.  Firstly, we derive a new class of exact inhomogeneous cosmological solutions whose matter content of the models is formed by a mixture of two interacting simple fluids plus a cosmological $\Lambda$-term. These models generalize the de Sitter
spacetime and the inhomogeneous two-fluid Szekeres-type cosmologies derived
by Lima and Tiomno. Finally, we show that the late time behaviour of our
solutions is in agreement with the ``cosmic no hair theorem'' of Hawking and
Moss.

\vspace*{.3cm}
PACS number(s): 04.20.Jb, 98.80.Dr
\end{abstract}

\thispagestyle{empty}
\maketitle
\section{Introduction}
\label{s1}
\def\theequation{\thesection.\arabic{equation}}
\setcounter{equation}{0}
It is widely known that inflationary models of the early universe provide a very interesting solution
to some shortcomings of the standard hot big-bang cosmology, among them the observed homogeneity, horizon and flatness problem \cite{GLA81,Lyth,Dod}. 
Although considering that many issues in inflationary cosmology remain unsettled, its confrontation 
with cosmic observations have already started with the latest cosmic microwave backgroun (CMB) experiments. Actually,  the three-year Wilkinson Microwave Anisotropy Probe (WMAP) data as well as the current five-year WMAP have precision enough to discriminate the simplest 
single-field inflationary models \cite{WMAP3,WMAP5}. Very recently, Kinney et al. \cite{Kiney} presented an update of the inflation
constraints using the WMAP5 and confirmed the results
previously obtained by themselves with the WMAP3
data. In particular, the WMAP5 team has claimed  that the quartic chaotic inflationary scenarios of
the form $V (\phi) = \lambda \phi^{4}$ were ruled out, while quadratic
chaotic inflationary models with potential, $V (\phi) = \lambda \phi^{4}$, revealed to be in better agreement
with the observational data \cite{WMAP5}.

In the simplest inflationary scenario, the initial conditions present in the big bang should be dynamically irrelevant to the structure of the Universe observed today. This  happen because during inflation the Universe expands almost exponentially, that is,  $a(t) \approx e^{Ht}$ ($H= \sqrt{\Lambda/3}$), where the cosmological constant $\Lambda$  represents the effective vacuum energy density of a given scalar field $\phi$. This can be taken for granted because in order to solve the cosmological puzzles the duration of such incredible expanding state must be at least of the order of 60 e-folds. The assumption that all expanding models with a positive cosmological constant evolve inevitably to the de Sitter spacetime means that the universe effectively loses the memory of its initial conditions.  Up to the present there is no general proof of this kind of ``cosmic no hair theorem'', as it was named by Hawking and Moss \cite{haw82}. Partial results
were obtained for non-rotating homogeneous and anisotropic cosmologies of the Bianchi type classification \cite{wald83}, as well as for some kind of
inhomogeneous Universes \cite{star83,jensen87}. Although the proof of the conjecture is only
shown in a restricted class of spacetimes, it is widely
believed that a large set of time-dependent cosmologies with $\Lambda$ must evolve to de Sitter spacetime which plays the hole of a special cosmic attractor, thereby potentially solving the dilemma of initial data.

On the other hand, some attention has also been paid to exact and approximated inhomogeneous models of
the universe in connection with the singularity problem as well as to solve
the difficulties in obtaining realistic inhomogeneous expanding solutions
endowed with a well defined equation of state \cite{Sen90,Rsen92,LN,Sussman2}. 
More recently, inhomogeneous cosmologies has also been considered as a possible 
explanation to the present accelerating stage of the Universe \cite{Moffat,PRD,CSL} 
with no appeal to some sort of dark energy component \cite{Review,decay}, or even to modified gravity theories \cite{Sotoriu,Santos}. 

In this framework, we derive here a large class of expanding inhomogeneous
solutions for which the equation of state problem is also circumvented in a natural way. These
solutions generalize a subclass of inhomogeneous models with a $\Lambda$-term
found by Barrow and Stein-Chabes \cite{jdb84}, and the inhomogeneous
two-fluid Szekeres-type cosmologies derived by Lima and Tiomno \cite{lt} (see also \cite{Sussman1}). In
Krasinski's  classification \cite{kra93}, such spacetimes are contained in
the subfamily of inhomogeneous solutions derived by Szafron \cite{Szafron77}, and, as we shall see, the complete set of 
solutions fulfill the requirements of the cosmic no hair conjecture. 
\section{The models}
\label{s2}
\def\theequation{\thesection.\arabic{equation}}
\setcounter{equation}{0}
Let us now consider the line element of the Szekeres-type models
of class II as
given in \cite{lt}(in our units $8\pi G=c=1$).
\begin{equation}
\label{e1}
ds^2 = dt^2 - Q^2\,dx^2 - R^2\, (dy^2 + h^2\,dz^2)\;,
\end{equation}
where

\begin{eqnarray}
\label{e2}
Q & \equiv &  Q(t,x,y,z) = R(t)\,A(x,y,z) +R_0\,T(t,x)\;,\\
\label{e3}
A(x,y,z) &=& 4\,\alpha(x)\,\left [\frac{\sin (\sqrt{k}\,y/2)}
{\sqrt{k}}\right ]^2 +\left [\chi(x)\,\cos z+\nu (x)\sin z\right ]\,
\frac{\sin (\sqrt{k}\,y)}{\sqrt{k}}+\nonumber\\
 &+& \xi(x) \cos(\sqrt{k}\,y)\;,
\end{eqnarray}

$\alpha$, $\chi$, $\nu$, $\xi$ are arbitrary functions of $x$,
$R_0$ is a constant factor with dimension of time, and
\begin{eqnarray}
\label{e4}
h(y) = \frac{\sin(\sqrt{k}\,y)}{\sqrt{k}} &=& y \hspace{1.15cm} {\rm for}
\hspace{.4cm} k=0\nonumber \\
 &=& \sin y \hspace{.6cm} {\rm for} \hspace{.4cm} k=+1\nonumber\\
 &=& \sinh y \hspace{.4cm} {\rm for} \hspace{.4cm} k=-1\;.
\end{eqnarray}
The functions $T(t,x)$ and $R(t)$ will be determined from the Einstein field
equations with a $\Lambda$-term,
\begin{equation}
\label{e5}
G^{\mu\nu} = T^{\mu\nu} +\Lambda g^{\mu\nu}\;,
\end{equation}
where the energy-momentum tensor $T^{\mu\nu}$ is given by
\begin{equation}
\label{e6}
T^{\mu\nu} = (\rho+p)u^{\mu}u^{\nu}-pg^{\mu\nu}\;,
\end{equation}
with $\rho$ and $p$ being the total energy density and pressure of the medium, respectively.

In the comoving frame, $u^{\mu}=\delta^{\mu}_0$, and it follows from
(\ref{e1})--(\ref{e6}) that the Einstein equations reduce to (a dot means derivatives with respect to the cosmic time)
\begin{eqnarray}
\label{e7}
\rho+\Lambda &=& \frac{3\, A\, R\,(\dot{R}^2+k)+ 2\,R_0\,R\,\dot{R}\,\dot{T}+
R_0\,T\,(\dot{R}^2+k)- 4\,\alpha\,R}{(A\,R +R_0\,T)\,R^2}\;,\\
\label{e8}
p-\Lambda &=& -2\frac{\ddot{R}}{R}-\frac{\dot{R}^2}{R^2}-\frac{k}{R^2}\;,\\
\label{e9}
\frac{2\alpha}{R_0} &=& R\ddot{T}+\dot{R}\dot{T}-\left [\ddot{R}+
\frac{\dot{R}^2+k}{R}\right ]\,T\;.
\end{eqnarray}

If $p=\Lambda=0$ the solutions of the above system take the form established by
Szekeres \cite{szek75,bonnor76}. If $p=0$ and $\Lambda\neq 0$ we get the
solutions found by Barrow and Stein-Schabes \cite{jdb84}. In the framework of a
two-fluid interpretation, the solutions derived by Lima and Tiomno \cite{lt}
have $\Lambda=0$ and
\begin{equation}
\label{e10}
\rho=\rho_{FRW}+\Delta\rho,
\end{equation}
where $\rho_{FRW}$ is a
Friedmannian homogeneous and isotropic component, and $\Delta\rho$ is an
inhomogeneous dust. In what follows we extend this approach by defining
\begin{eqnarray}
\label{e11}
\rho_{FRW} +\Lambda\ &\equiv& 3\,\frac{\dot{R}^2+k}{R^2}\;,\\
\label{e12}
\Delta\rho &\equiv& \frac{2\,R_0\,R\,\dot{R}\,\dot{T} - 2\,R_0\, T\,(\dot{R}^2+
k)-4\,\alpha\, R}
{(A\, R +R_0\,T)\,R^2}\;.
\end{eqnarray}
If we now take
\begin{equation}
\label{e13} p=\omega\rho_{FRW}\;,
\end{equation}
equation (\ref{e8}) can be recast as
\begin{equation}
\label{e14} R\,\ddot{R} +\left (\frac{1+3\omega}{2}\right
)\,(\dot{R}^2+k)-\frac{1}{2} {(1+\omega)}\,\Lambda\,R^2=0\;,
\end{equation}
whose first integral is given by
\begin{equation}
\label{e15} \dot{R}^2=\left (\frac{R_0}{R}\right
)^{1+3\omega}-k+\frac{1}{3}\,\Lambda\,R^2 \;.
\end{equation}
We note that eq. (\ref{e14}) is the general FRW differential equation for a fluid
with a gamma-law equation of state plus a $\Lambda$-term. Also, the functional
dependence of the Friedmanian component $\rho_{FRW}$ with respect to the scale
factor is obtained by replacing the first integral (\ref{e15}) into (\ref{e11}).
Moreover, inserting (\ref{e14}) into (\ref{e9}), we see that
the equation for $T(t,x)$ becomes
\begin{equation}
\label{e16} R\,\ddot{T} +\dot{R}\,\dot{T}+\left
[\frac{1-3\omega}{1+3\omega}\,\ddot{R}
-\frac{1+\omega}{1+3\omega}\,\Lambda\,R\right
]\,T=\frac{2\alpha}{R_0}\;.
\end{equation}
The problem now reduces to solve the equations (\ref{e14}) or
equivalently (\ref{e15}) and (\ref{e16}).  Following the procedure
developed in Refs. \cite{lt,LN} we find that, if $k=\alpha=0$
and $\omega\neq-1$, the functions $R$ and $T$ can be written as

\begin{eqnarray}
\label{e17}
R &=& R_0\,\left [\cosh\Theta(t)+\frac{\sqrt{3/\Lambda}}{R_0}\,\sinh\Theta(t)
\right ]^{\frac{2}{3(1+\omega)}}\;,\\
\label{e17.5}
\Theta(t) &=& \frac{\sqrt{3\Lambda}}{2}\,(1+\omega)\,(t-t_0)\;,\\
\label{e18} T &=& \beta (x)\,\left (\frac{R}{R_0}\right )+\mu
(x)\,\left (\frac{R}{R_0}\right )^{ \frac{-1+3\omega}{2}}\:F\left
[\frac{\omega-1}{2(1+\omega)},\frac{1}{2},\,\frac{1+3
\omega}{2(1+\omega)};\, z\right ]\;,\\
\label{e18.1} z &=& -\frac{1}{3}\,\Lambda\,R_0^2\left
(\frac{R}{R_0}\right ) ^{3(1+\omega)}\;,
\end{eqnarray}
where $\beta$ and $\mu$ are two new arbitrary functions of $x$ and $F$ is the
hypergeometric Gaussian function.

For $k=\alpha=0$ and $\omega=-1$ the solutions an be written as 
\begin{eqnarray}
\label{e18.2}
R &=&R_0\,\exp \left [\sqrt{\frac{1}{R_0^2}+\frac{\Lambda}{3}(t-t_0)}\,\right ]
\;,\\
\label{e18.3}
T &=& \beta (x)\,\left (\frac{R}{R_0}\right ) + \mu (x)\,\left
(\frac{R}{R_0}\right )^{-2}\;.
\end{eqnarray}

In order to complete the solution it is sufficient to substitute the above
equations in the expressions (\ref{e11}) and (\ref{e12}) for the density. As
one may check, these solutions have no killing vectors, are type D in the
Petrov classification, and the flow of matter is irrotational and geodetic.

As one should expect, if $\Lambda=0$, (\ref{e17}) -- (\ref{e18.1}) reduce to
\begin{equation}
\label{e19} R=R_0\,\left
[1+\frac{3(1+\omega)}{2}\,\frac{(t-t_0)}{R_0}\right ]^{\frac{2}
{3(1+\omega)}}\;,
\end{equation}
and
\begin{equation}
\label{e20} T= \beta (x)\,\left (\frac{R}{R_0}\right ) + \mu
(x)\,\left (\frac{R}{R_0} \right )^{\frac{-1+3\omega}{2}}\;,
\end{equation}
which are the expressions derived in ref. \cite{lt} for
$k=\alpha=0$ and $\omega\neq -1$. Furthemore, using the identity
$F(a,b,b;z)=(1-z)^{-a}$ \cite{abr}, it is easy to see that, if
$\omega=0$ and $\Lambda\neq 0$, a subclass of the parabolic models
derived by Barrow and Stein-Chabes is recovered.

For large values of the cosmological time, we have
\begin{eqnarray}
\label{e21}
R &\sim & e^{\sqrt{\Lambda /3}\,t}\;,\\
\label{e22}
T &\sim & \beta (x)\, \left (\frac{R}{R_0}\right )\;.
\end{eqnarray}
In this limit one can see that $\sigma \sim 0$ and $\theta\sim\sqrt{3\Lambda}$,
where $\sigma$ and $\theta$ are the shear and expansion rates, respectively.
Hence, since $Q\sim A\,R$, the homogeneous and isotropic de Sitter phase is
reached (compare with the results of Lima and Tiomno \cite{lt}).
As in the solutions presented by Barrow and Stein-Schabes \cite{jdb84}, the
comoving geodetic observers of our models will see the deviations of the
de Sitter exact form decay exponentially with time in agreement with the
``cosmic no hair conjecture''.

Finally, we remark that the existence of this large class of exact inhomogeneous cosmological  models evolving to de Sitter spacetime may have great interest for inflationary scenarios. In particular, for  the whole class of Szekeres type cosmologies, which includes the closed and open FRW universes as limiting case (Lima and Tiomno{\cite{lt}}), one may investigate how the flatness problem is solved under more general conditions than the ones usually considered in the inflationary context.
\section*{Acknowledgments}

The authors would like to thank  CNPq and FAPESP (Brazilian Research Agencies) for
financial support.

\end{document}